\documentclass[printer]{aa}
\usepackage{graphicx}
\usepackage{multirow}
\usepackage{natbib}
\bibpunct{(}{)}{;}{a}{}{,} 
\usepackage[varg]{txfonts}

\def\hdsqt{\mbox{H\,1743-322}}
\def\swiftjdsqc{\mbox{Swift\,J1745-26}}
\def\swiftjdsct{\mbox{Swift\,J1753.5-0127}}
\def\xtejodh{\mbox{XTE J1118+480}}
\def\xtejqcq{\mbox{XTE J1550-564}}
\def\Av{A_{\rm V}}
\def\mags{\mbox{ magnitudes}}
\def\kpc{\mbox{ kpc}}
\def\ergs{\mbox{ erg\,s}^{-1}}
\def\hour{^{h}}
\def\min{^{m}}
\def\asec{^{\prime \prime}}
\def\secp{{\rlap.}^{s}}
\def\adeg{^{\circ}}
\def\amin{^\prime}
\def\asecp{{\rlap.}^{\prime \prime}}
\def\micron{ \mu \mbox{m}}
\def\ergcms{\mbox{ erg s}^{-1} \mbox{cm}^{-2}}
\def\nh{N_{\rm H}}
\def\cmmoinsdeux{\mbox{ cm}^{-2}}
\def\Msol{\mbox{ }M_{\odot}}
\def\Rsol{\mbox{ }R_{\odot}}
\def\Hz{\mbox{ Hz}}
\def\K{\mbox{ K}}

\begin{document}

\title{Broad-band spectral energy distribution of the X-ray transient $\swiftjdsqc$ from outburst to quiescence\thanks{Based on observations collected at the European Organisation for Astronomical Research in the Southern Hemisphere under ESO programmes 089.D-0191 and 090.D-077 (PI Chaty).}}

\author{Sylvain Chaty\inst{1,2,3}
  \and Francis Fortin\inst{1,2,3}
  \and Alicia L\'opez-Oramas\inst{4}}

\authorrunning{Chaty et al.}
\titlerunning{OIR observations of the black hole X-ray transient \swiftjdsqc}

 \institute{Universit\'e de Paris, AIM, F-91191 Gif-sur-Yvette, France
   \and AIM, CEA, CNRS, Universit\'e Paris-Saclay, F-91191 Gif-sur-Yvette, France
   \and Universit\'e de Paris, CNRS, APC, F-75013 Paris, France
  \and Instituto de Astrof\'isica de Canarias, Universidad de La Laguna, Dpto. Astrof\'isica, E-38206 La Laguna, Tenerife, Spain  
}

\date{Received xxx / Accepted xxx}

\abstract {}
          {We aim to analyse our study of the X-ray transient $\swiftjdsqc$, using observations obtained from its outburst in September 2012, up to its decay towards quiescence in March 2013.}
          {We obtained optical and infrared observations, through override programme at ESO/VLT with FORS2 and ISAAC instruments, and added archival optical (VLT/VIRCAM), radio and X-ray ({\textit Swift}) observations, to build the light curve and the broad-band spectral energy distribution (SED) of $\swiftjdsqc$.}
          {We show that, during its outburst and also during its decay towards quiescence, $\swiftjdsqc$ SED can be adjusted, from infrared up to X-rays, by the sum of both a viscous irradiated multi-colour black body emitted by an accretion disc, and a synchrotron power law at high energy. In the radio domain, the SED arises from synchrotron emission from the jet. While our SED fitting confirms that the source remained in the low/hard state during its outburst, we determine an X-ray spectral break at frequency $3.1 \leq \nu_{break} \leq 3.4 \times 10^{14}$\,Hz, and a radio spectral break at $10^{12}$\,Hz $\leq \nu_{break} \leq 10^{13}$\,Hz. We also show that the system is compatible with an absorption $\Av$ of $\sim 7.69 \mags$, lies within a distance interval of D$ \sim[2.6-4.8] \kpc$ with an upper limit of orbital period P$_{orb} = 11.3$~hours, and that the companion star is a late spectral type in the range K0 -- M0\,V, confirming that the system is a low-mass X-ray binary. We finally plot the position of $\swiftjdsqc$ on an optical-infrared -- X-ray luminosity diagram: its localisation on this diagram is consistent with the source staying in the low-hard state during outburst and decay phases.}
        {By using new observations obtained at ESO/VLT with FORS2 and ISAAC, and adding archival optical (VLT/VIRCAM), radio and X-ray ({\textit Swift}) observations, we built the light curve and the broad-band SED of $\swiftjdsqc$, and we plotted its position on an optical-infrared -- X-ray luminosity diagram. By fitting the SED, we characterized the emission of the source from infrared, via optical, up to X-ray domain, we determined the position of both the radio and X-ray spectral breaks, we confirmed that it remained in the low-hard state during outburst and decay phases, and we derived its absorption, distance interval, orbital period upper limit, and the late-type nature of companion star, confirming $\swiftjdsqc$ is a low-mass X-ray binary.}

\maketitle

\section{Introduction}

Soft X-ray transients (SXT) are low-mass X-ray binary systems (LMXB), composed of a low-mass (M $\le$ 1 M$_{\odot}$) star and a compact object, either a neutron star or a black hole \citep[see][]{chaty:2013}. These systems spend most of their lives in quiescence (typically years to decades) when their X-ray luminosity is very low \citep[L$_{X} \sim 10^{31} \ergs$, i.e. well below the Eddington luminosity;][]{gallo:2012}. From time to time, they exhibit violent outbursts at all wavelengths, which evolve on short timescales of days (or even less). During these outbursts, matter from the star is accreted onto the compact object, via Roche-Lobe overflow and accumulation in an accretion disc, reaching bright X-ray luminosities up to L$_{X} \sim 10^{38-39} \ergs$. The behaviour of SXT in both radio and X-rays is well described by a unified model \citep{fender:2010}. A typical outburst is characterised by a fast rise in luminosity, with hard X-ray emission dominated by a hot inner accretion disc and synchrotron emission coming from a radio jet (low/hard state), followed by an exponential decay, until the stellar emission begins to be revealed in the optical and infrared (OIR), and eventually the jet disappears (high/soft state).

One such source is the transient black-hole candidate Swift J174510.8-262411 (\swiftjdsqc\ hereinafter), discovered by {\textit Swift}/BAT (15 -- 50 keV) on September 16, 2012 \citep[MJD 56186.39,][]{cummings:2012a} and detected by {\textit Swift}/XRT \citep[0.2 -- 10 keV,][]{cummings:2012b}, located at RA (J2000) = $17\hour 45\min 10\secp82$, Dec (J2000) = $-26\adeg 24\amin 12\asecp7$. The flux increase was similar to those observed in black-hole transients. It was classified as a failed transition, since it did not reach the soft state \citep{belloni:2012}. Following the outburst rise, the source remained in the hard state \citep{tomsick:2012c,sbarufatti:2013}, and it changed to hard/intermediate state in MJD 56370 \citep{belloni:2012}, showing a slightly steeper spectrum. A secondary flare was detected in optical and X-rays on MJD 56380  \citep{russell:2013b}. The outburst remained observable until June 2013. Follow-up observations in the optical confirmed the presence of a double-peaked H$\alpha$ line, characteristic of black-hole transients \citep{deugartepostigo:2012,tomsick:2012c}. It was proposed that $\swiftjdsqc$ be classified as an LMXB, with a companion star of spectral type later than A0 and orbital period of less than 21\,hours \citep{munoz-darias:2013}. Radio and sub-millimetre (sub-mm) observations performed with VLA, SMA, and JCMT in September 2012 revealed a power law with inverted spectral index $\alpha \sim 0.07-0.17$ \citep{tetarenko:2015}. The outburst decay was well studied in optical and X-ray wavelengths by \cite{kalemci:2014}, who concluded that the X-ray spectra were consistent with thermal Comptonization, without ruling out a jet synchrotron origin with a high-energy cutoff at 112\,keV.

We triggered observations of $\swiftjdsqc$ at ESO at two different epochs, the first one during its outburst rise in September 2012, and the second one in March 2013 as soon as it was observable again, corresponding to its decay towards quiescence. In the following work, we present our study of $\swiftjdsqc$: we describe the observations in Section \ref{section:observations}, the results in Section \ref{section:results}, the discussion in Section \ref{section:discussion}, and the outcomes in Section \ref{section:conclusion}.

\section{Observations\\ } \label{section:observations}

{\bf We now describe the observations we performed on the source $\swiftjdsqc$.}

\subsection{VLT/FORS2 and ISAAC photometric observations}

We triggered two sets of observations of $\swiftjdsqc$: firstly from September 18, 2012 (ESO ID 089.D-0191, PI Chaty, two nights after the {\textit Swift}/BAT discovery alert), lasting (with uneven sampling) until September 25, 2012; and the second from March 06 to 13, 2013 (ESO ID 090.D-077, PI Chaty), covering the decay towards quiescence.

We performed observations with Unit Telescope 1 (UT1) of the Very Large Telescope (VLT) in the OIR domain. Optical observations were carried out with the FORS2 instrument, covering the $0.33-1.10 \micron$ band using U (identified as u-HIGH, $\lambda = 0.365 \micron$), B (b-HIGH, $\lambda = 0.437 \micron$), V (v-HIGH, $\lambda = 0.555 \micron$), R (R-special, $\lambda = 0.655 \micron$), and I (I-BESS, $\lambda = 0.768 \micron$) filters. For the infrared observations we used the ISAAC instrument, covering the band between $1-2.5 \micron$, using the filters J$_{s}$ ($\lambda = 1.240 \micron$), H ($\lambda = 1.664 \micron$), and K$_{s}$ ($\lambda = 2.164 \micron$). Air masses were always between 1.03 and 1.40.

{\bf All optical and infrared} data were reduced using standard IRAF routines, with bias and flat-field correction \citep{tody:1986,tody:1993}. In addition, the contribution from the thermal sky emission was removed from the infrared data, by pointing at different positions in the sky, that we combined through median-filter to produce a blank thermal sky, which was then removed from our data. The obtained fluxes were calibrated with those of standard stars: PG2213-B and E7-S6 for September 2012, FS121 and E7-S6 for March 2013.

\begin{figure}
\centering
\includegraphics[width=.51\textwidth]{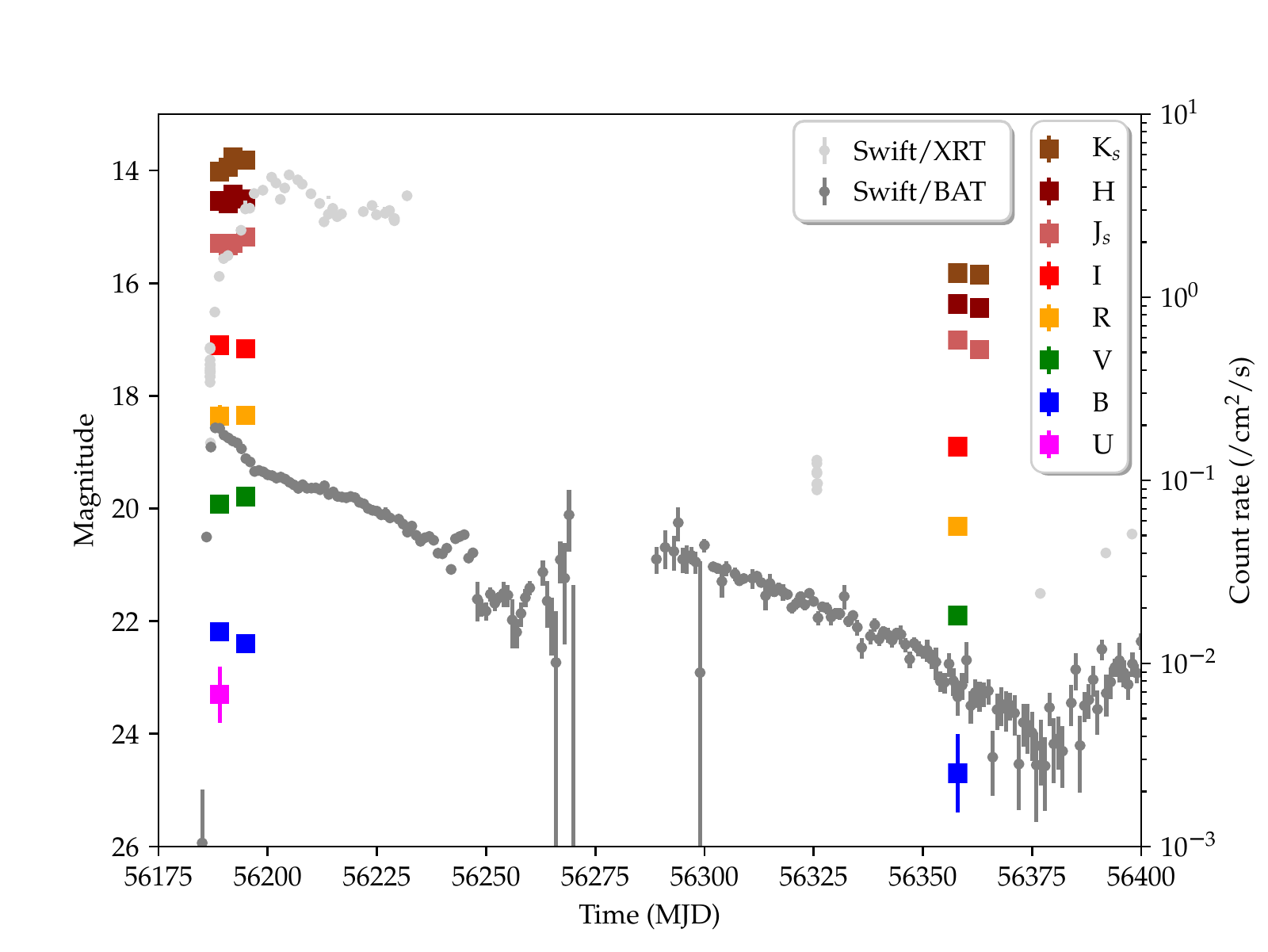}
\caption{Optical, infrared, and X-ray light curves of $\swiftjdsqc$ during its 2012-2013 outburst. UBVRI and J$_s$HK$_s$ photometry come from the VLT/FORS2 and ISAAC observations, respectively. The high-energy {\textit Swift}/BAT light curve comes from archival data. \label{fig:lc}}
\end{figure}

In Table~\ref{table:mag}, we report our photometry results for both data sets (September 2012 and March 2013, respectively). The OIR light curve is shown in Figure~\ref{fig:lc}, overplotted with the X-ray {\textit Swift} light curve in order to show the state of the source during its outburst rise, and its decline towards quiescence, respectively. The general trend shows that both OIR and X-ray fluxes decrease from September 2012 to March 2013, indicating that $\swiftjdsqc$ exhibits high activity during the first epoch, decaying towards quiescence during the second epoch.

\begin{table*}
  \scriptsize
  \begin{tabular}{cccccccccc}
    \hline
    Date & MJD & U($0.361 \micron$) & B($0.437 \micron$) & V($0.555 \micron$) & R($0.655 \micron$) & I($0.768 \micron$) & J$_S$($1.240 \micron$) & H($1.664 \micron$) & K$_S$($2.164 \micron$)\\
 & Exp time (s) & $60 \times 3$ & $60 \times 3$ & $60 \times 3$ & $60 \times 3$ & $60 \times 3$ & $30 \times 9$ & $10 \times 9$ & $10 \times 9$ \\
    \hline
    %
    18/09/2012 & 56189 & 23.3$\pm$0.5       & 22.19$\pm$0.07   & 19.92$\pm$0.08   & 18.4$\pm$0.2    & 17.1$\pm$0.1     & 15.29$\pm$0.06 & 14.54$\pm$0.06 & 14.02$\pm$0.03 \\
    20/09/2012 & 56191 & --                & --              & --              & --              & --              & 15.33$\pm$0.07 & 14.59$\pm$0.06 & 13.94$\pm$0.03 \\
    22/09/2012 & 56192 & --                & --              & --              & --              & --              & 15.29$\pm$0.07 & 14.42$\pm$0.05 & 13.76$\pm$0.03 \\
    25/09/2012 & 56195 & --                & 22.40$\pm$0.1    & 19.79$\pm$0.04   & 18.35$\pm$0.04   & 17.16$\pm$0.06   & 15.18$\pm$0.07 & 14.51$\pm$0.06 & 13.82$\pm$0.03 \\
    \hline\\[-1.5ex]
    06/03/2013 & 56358 & --                & 24.7$\pm$0.7     & 21.9$\pm$0.1     & 20.32$\pm$0.04   & 18.90$\pm$0.09   & 17.01$\pm$0.04 & 16.37$\pm$0.05 & 15.82$\pm$0.06 \\
    12/03/2013 & 56363 & --                & --              & --              & --              & --              & 17.18$\pm$0.05 & 16.44$\pm$0.06 & 15.85$\pm$0.04 \\
    \hline
  \end{tabular}
\footnotesize
 \caption{OIR photometry results of $\swiftjdsqc$ during the 2012-2013 outburst, with the date (days, MJD) and apparent magnitudes for the different optical (u, b, v, R and I) and infrared (J$_{s}$, H and K$_{s}$) filters.}\label{table:mag}
\end{table*}

\subsection{VLT/FORS2 spectroscopic observations}

We also obtained 340--620\,nm optical spectra of $\swiftjdsqc$ with the FORS2 instrument on September 19, 2012, and March 06, 2013 (exposure time: 10 min, slit: $0\asecp7$, grism: GRIS\_600B, instrument mode: FORS2\_lss\_obs\_off\_fast). However, both spectra do not show any discernable spectral feature, so most likely emanate from the accretion disc emission, and/or a non-thermal synchrotron component.

\subsection{VLT/VIRCAM observations}

\begin{figure}
\centering
\includegraphics[width=.49\textwidth]{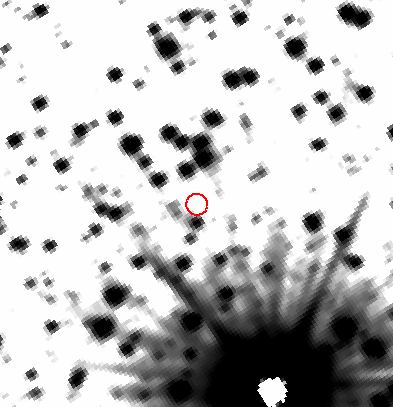}
\caption{VIRCAM $38\asec \times 38\asec$ field of view, North to the top, East to the left, with a $1\asec$ circle around the position of $\swiftjdsqc$. \label{fig:vircam}}
\end{figure}

We retrieved a $12 \times 4$\,s Ks band image acquired by ESO/VIRCAM on April 9, 2019 of the field around $\swiftjdsqc$. The acquisition was performed under the VVV Extended ESO Public Survey (VVVX, programme \# 198.B-2004(I)). Due to the tiling of the VIRCAM detector, the effective exposure time on the field around the source is $4 \times 4 = 16$\,s. We do not detect the source at this epoch. The limiting Ks magnitude taken at 5$\sigma$ above the background noise was computed at K$_s = 18.2 \mbox{ magnitude}$ (in the Vega system). As the source is in quiescence, we suggest that K$_s(lim) = 18.2 \mbox{ magnitude}$ can be taken as a lower limit for the apparent magnitude of the companion star. We show the VIRCAM $38\asec \times 38\asec$ field of view around the position of $\swiftjdsqc$ in Figure~\ref{fig:vircam}.

\section{Results} \label{section:results}

\begin{figure}
\centering
\includegraphics[width=.51\textwidth]{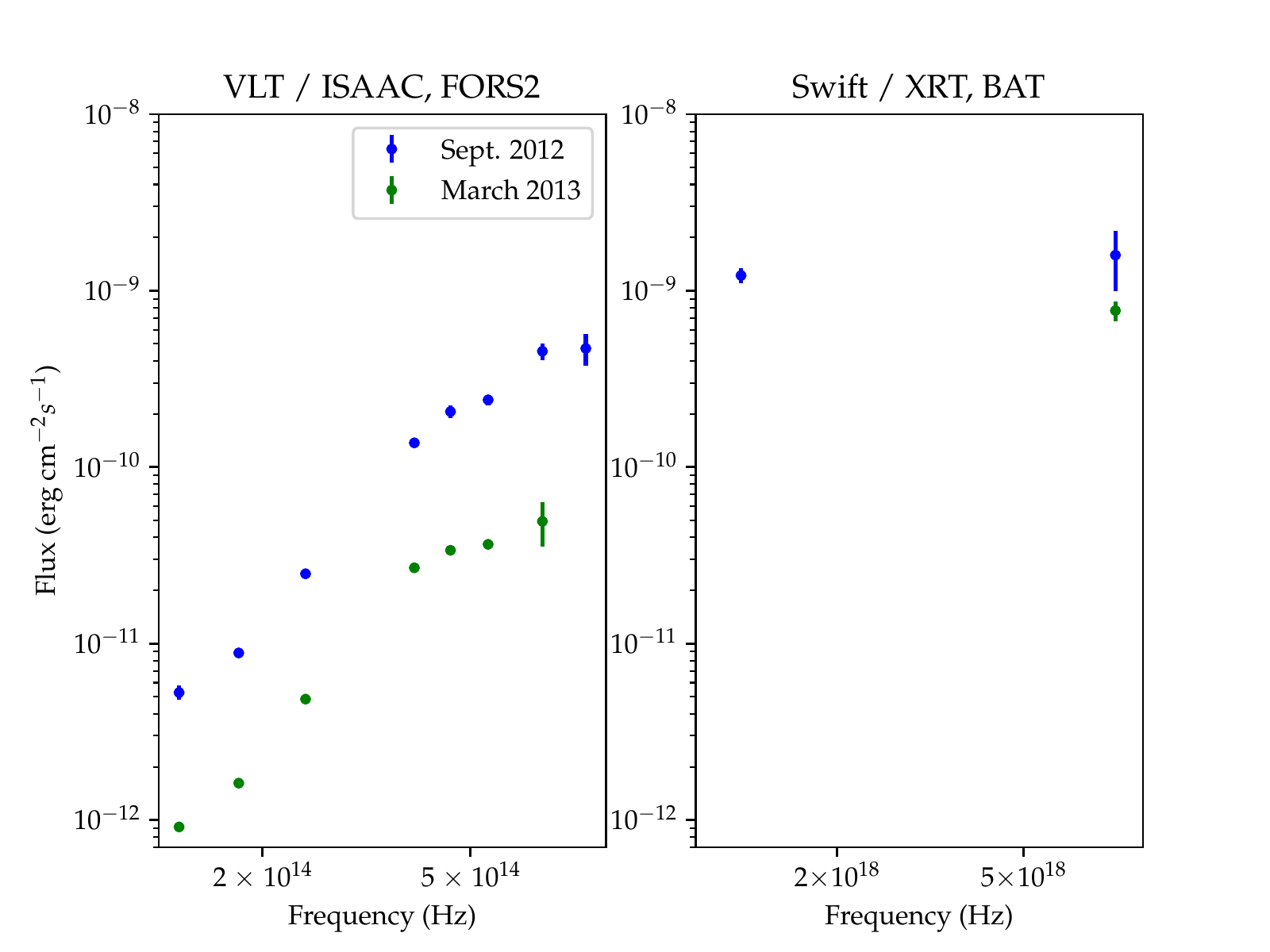}
\caption{Spectral energy distribution of VLT/ISAAC (infrared), FORS2 (optical) and {\textit Swift}/XRT, BAT (soft and hard X-rays) data points of $\swiftjdsqc$ during its 2012-2013 outburst. \label{fig:sed}}
\end{figure}

We use our OIR (VLT/FORS2 and ISAAC) measurements of September 2012 and March 2013, VIRCAM upper limit, and archival X-ray observations obtained at similar epochs by {\textit Swift}/BAT \citep{sbarufatti:2013,kalemci:2014} and {\textit Swift}/XRT \citep{sbarufatti:2012}, to build an OIR-to-X-ray spectral energy distribution (SED) of \swiftjdsqc, as shown in Figure~\ref{fig:sed} (OIR in left panel and X-rays in right panel, respectively).

\subsection{Modelling the spectral energy distribution}

We fit this SED for both epochs independently. $\swiftjdsqc$ being a (likely black hole) low-mass X-ray binary, we expect that its OIR thermal emission comes from the sum of an accretion disc and a late-type main sequence star. The high-energy emission detected by {\textit Swift} likely arises from a corona, or from the base of a jet, around the compact object.
We thus use three different contributions in our model: firstly, a multi-colour black-body acccretion disc; secondly, a cut-off power law for the black-hole corona and base of jet; and, thirdly, a standard black-body spectrum for the low-mass late-type stellar companion.

We adjust the OIR SED of both epochs with a viscous-irradiated disc model \citep{hynes:2002a}, meaning a sum of a viscous multi-colour black-body spectrum \citep{shakura:1973}, with the corresponding modified temperature distribution of an irradiated disc \citep{cunningham:1976}.
The multi-colour black-body component is thus modeled as follows:
\begin{equation}
F_{\nu} = 2\pi cos(i) D^2 \int_{R_{in}}^{R_{out}}rB_{\nu}\left(T(r)\right)dr ,
\end{equation}

with B$_{\nu}$ the Planck function for black-body radiation\,:
\begin{equation}
B_{\nu}\left(T(r)\right) = \frac{2h\nu^3}{c^2}\frac{1}{exp\left(\frac{h\nu}{kT(r)}\right)-1} ,
\end{equation}

and T(r) the effective temperature at radius $r$:
\begin{equation}
T^4(r) = T_{v}^4(r) + T_{i}^4(r) .
\end{equation}

The temperature profiles due to viscous heating and irradiation are, respectively, defined as:
\begin{equation}
T_{v}(r) = T_{v}\left(\frac{r}{R_{out}}\right)^{-\frac{3}{4}}
\mbox{and }
T_{i}(r) = T_{i}\left(\frac{r}{R_{out}}\right)^{-\frac{3}{7}} .
\end{equation}

This multi-colour black-body model uses six parameters: the inner radius R$_{in}$ and outer radius R$_{out}$ of the accretion disc, the viscous temperature T$_v$, the irradiated temperature T$_i$, the inclination angle $i$, and the distance D.

Concerning the high-energy emission, the X-ray spectrum is adjusted with a power law, defined by two parameters --the amplitude $A$ and the spectral index $E$-- to which we add a low-energy exponential cutoff at the frequency $\nu_0$, with the exponential decrease rate governed by the index $P$, influencing the speed at which the exponential cutoff happens. The equation for the power law (Eq. 5), governed by the four parameters \{A,~E,~$\nu_0$,~P\}, is written as:

\begin{equation}
Pl_{\nu} = \frac{A\times F_A\times\left(\frac{\nu_0}{\nu_A}\right)^P\times\left(\frac{\nu}{\nu_A}\right)^{E-P}}{exp\left(\left(\frac{\nu}{\nu_0}\right)^{-P}\right)-1} .
\end{equation}

The amplitude $A$ is normalised using the flux F$_A$ and frequency $\nu_A$ of the lowest energy {\textit Swift} data point of 2012, September ($1.22 \times 10^{-9} \ergcms$ flux at $1.25 \times 10^{18}$\,Hz). This presents the advantage of having the amplitude $A$ normalised, independently of the spectral index $E$ and the exponential decrease rate $P$. Using this equation, the cutoff power-law value at $\nu_0$ is $\sim 1.7$ times lower than the equivalent classical power law.

\subsection{Extinction}

\begin{figure}
\centering
\includegraphics[width=.54\textwidth]{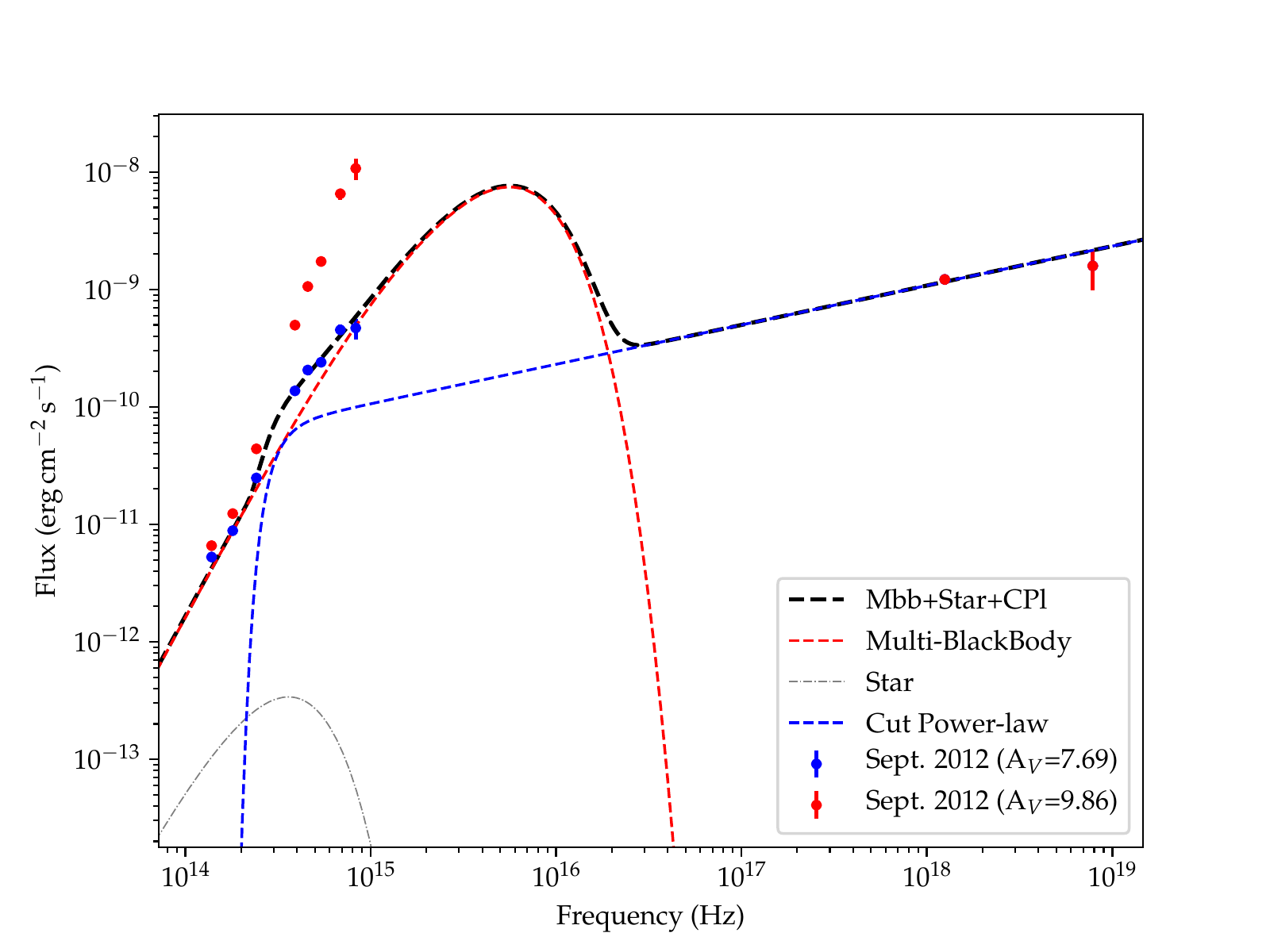}
\caption{SED of $\swiftjdsqc$ from OIR to X-rays for September 2012 observations. The SED is built with data points de-reddened with an $\Av = 7.69$ (blue points) and $\Av = 9.86 \mags$ (red points), respectively. The result of the fit is plotted in black, adding the stellar black body (grey), the multi-colour black-body accretion disc (red), and the X-ray power law (blue). While it is impossible to fit the optical data points de-reddened with an A$_{V}$ of $9.86$, a reasonable fit can be obtained with optical and infrared data points de-reddened with an $\Av$ of $7.69 \mags$.
  \label{fig:sed-ext}}
\end{figure}

In order to correct OIR magnitudes for interstellar reddening, we use the column density of $\swiftjdsqc$, as measured in X-rays. \cite{tomsick:2012c} derived a value of $\nh = (1.70\pm0.04) \times 10^{22} \cmmoinsdeux$ (on MJD 56188, during the rise of the outburst), while \cite{kalemci:2014} obtained $\nh = (2.18 \pm 0.25) \times 10 ^{22} \cmmoinsdeux$ (by fitting X-ray spectra taken for several nights from MJD 56334 to MJD 56435, during the decay). In this paper, we use the relationship between hydrogen column density --N$_{H}$-- and optical extinction --A$_{V}$-- in our Galaxy given by \cite{guver:2009}, based on pointed X-ray observations of a large sample of supernova remnants. We derive a total extinction of the system, assuming the two values of N$_{H}$, respectively, of $\Av = 7.69$ and $9.86$ magnitudes. In Fig. \ref{fig:sed-ext}, we show the results of our fit for these two values of extinction. By taking the high level of absorption (A$_{V}$ = 9.86), the optical data are too bright to be fitted with both a multi-colour black-body disc and an X-ray power law. On the other hand, when choosing the low level of absorption ($\Av = 7.69$), we can reasonably fit both the infrared and optical data by summing the multi-colour black-body disc with the X-ray power law. In the following, we therefore use the low level of absorption ($\Av = 7.69$) to correct the data for interstellar redening. Considering R$_{V} = A_{V} / E(B-V) = 3.1$, we obtain the extinction coefficients for the other wavelengths \citep{cardelli:1989}: A$_B =10.28$, A$_R = 5.75$, A$_I = 3.68$, A$_{J_{s}}= 2.17$, A$_{H}= 1.46$ and A$_{K_{s}}=0.88$ magnitudes.

\subsection{Distance}

The distance can not be derived directly from SED fitting, thus we used the VIRCAM non-detection as an upper limit of the flux of $\swiftjdsqc$ in quiescence, to derive the confidence interval on the distance for a black body corresponding to a companion star of a certain spectral type. We thus performed some preliminary fits to explore the space parameters, by interpolating the spectral type and the distance obtained, using the VIRCAM lower limit as an input. We find that spectral types earlier than K0, corresponding to distances higher than 4.8 kpc, give us disc temperatures higher than T$\geq 25 000$K, which we reject based on the highest disc temperature given in \cite{hynes:2002a}. Similarly, spectral types later than M0, located closer than 2.6 kpc, give us disc temperatures lower than T$\leq 1400$K, which we also reject based on the lowest disc temperature in \cite{hynes:2002a}.
We thus find a valid distance interval D of the system comprised between 2.6 and 4.8 kpc, corresponding to a stellar spectral type interval of the companion star comprised between K0 and M0\,V.

\subsection{Orbital period and separation}

From our photometric measurements, we can set further constraints on the orbital period of the system, following the method by \cite{munoz-darias:2013}. The apparent magnitude in the R band during the outburst is R$_{outburst} = 18.4 \pm 0.2$ on MJD 56189. \cite{hynes:2012} determined that the magnitude in quiescence was R$_{quiescence} \geq 23.1 \pm 0.5 \mags$. 
Taking the spectral type interval of [K0-M0] derived earlier, and the extinction, along with the K$_s(lim) = 18.2 \mbox{ magnitude}$ obtained from VIRCAM observations, we derive a new upper limit at R$_{quiescence} \geq 25.0 \mbox{ magnitude}$, more constraining than the one given by \cite{hynes:2012}.
Thus, we obtain $\Delta R = R_{quiescence} - R_{outburst} \geq 6.6 \pm 0.2 \mbox{ magnitude}$. From Table~\ref{table:mag}, we assume that $\Delta R \sim \Delta$V, since the trend is similar and the spectrum is disc dominated. We apply formula~(1) from \cite{shahbaz:1998}, which provides an empirical relationship between the outburst amplitude and the orbital period of a system. We obtain an orbital period upper limit P$_{orb} \leq 11.3 $\,hours, more constraining than the upper limit of P$_{orb} = 21$\,hours proposed by \cite{munoz-darias:2013}. Considering Formula~(2) from \cite{faulkner:1972}, the obtained orbital period upper limit implies an average density for the companion star of $\rho > 1.21$\,g.cm$^{-3}$, corresponding to a spectral type F4\,V or later \citep{kreiken:1953}, in agreement with \cite{munoz-darias:2013}, and consistent with the [K0-M0] spectral type interval. Hence, we confirm that $\swiftjdsqc$ is an LMXB.


Finally, we need to know the size of the accretion disc, constrained between the inner radius R$_{in}$ and the outer radius R$_{out}$.
Assuming a mass of the (likely) black hole of $10\Msol$, we will fit R$_{in}$, defined as a multiple of the innermost stable circular orbit (ISCO):
\begin{equation}
ISCO = \frac{6\,GM}{c^2} = 3\,R_{S} \sim 1.3\times10^{-4} \Rsol .
\end{equation}

With a Roche lobe around the black hole equal to R$_L = 3.1 \Rsol$ \citep{eggleton:1983}, we derive R$_{out} = 0.9 \times R_L = 2.78\Rsol$. Using the orbital period upper limit obtained above, we compute, assuming a circular orbit, an orbital separation of a~$ \leq 5.1 \pm 1.9 \Rsol$. Finally, we assume a fixed inclination angle of i$ = 45 \adeg$.

In the following, we present the results of our fits performed for a K5\,V star (R$ = 0.72 \Rsol$, M$ = 0.67 \Msol$, T$_{eff} = 4410$K, \citeauthor{allen:1973} \citeyear{allen:1973}), right in the middle of the [K0 -- M0] spectral range, located at a distance of D$ = 3.5 \kpc$.


\begin{table*}
\small
\begin{tabular}{lc|ccc|cccc}
\hline\\
      &               & \multicolumn{3}{c}{Multi-black body}  & \multicolumn{4}{c}{X-ray power law} \\
Epoch & $\chi^2_{\nu}$ (dof) & R$_{in}$    & T$_{v}$  & T$_{i}$  & $E$ & $A$ & $\nu_0$ & $P$ \\
      &               & (ISCO)      & (K) & (K) &   & ($1.22 \times 10^{-9}\ergcms$) & ($10^{14}\Hz$) & \\
\hline\\[-1.5ex]
Sept. 2012 & $1.923$ (4) & 300\,(f)    & 0(f) & $16400\pm700$ & $0.32\pm0.05$ & $1.0\pm0.1$ & $3.4\pm0.9$ & $4.4\pm3.2$ \\
           & $2.029$ (4) & $1600$\,(f) & $13500\pm600$ & 0(f) & $0.34\pm0.05$ & $1.0\pm0.1$ & $3.2\pm0.9$ & $5.0\pm4.5$ \\
\hline
March 2013 & $1.743$ (3) & $1300$\,(f) & 0 (f) & $4950\pm70$ & $0.34\pm0.02$ & $0.34\pm0.05$ & $3.3\pm0.3$ & $4.4\pm1.1$ \\
           & $2.400$ (3) & $2900$\,(f) & $4240\pm70$ & 0(f) & $0.35\pm0.02$ & $0.33\pm0.05$ & $3.2\pm0.3$ & $4.9\pm1.7$ \\
\hline\\
\end{tabular}
\normalsize
\caption{Parameters obtained from fitting the September 2012 and March 2013 SED of $\swiftjdsqc$, including ISAAC, FORS2 and {\textit Swift} data (`f' means frozen). The quoted errors are obtained for $\Delta \chi^2 = 1.0$, corresponding to a 68\% confidence limit.
}\label{table:fit-results}
\end{table*}

\begin{figure}
\centering
\includegraphics[width=0.54\textwidth]{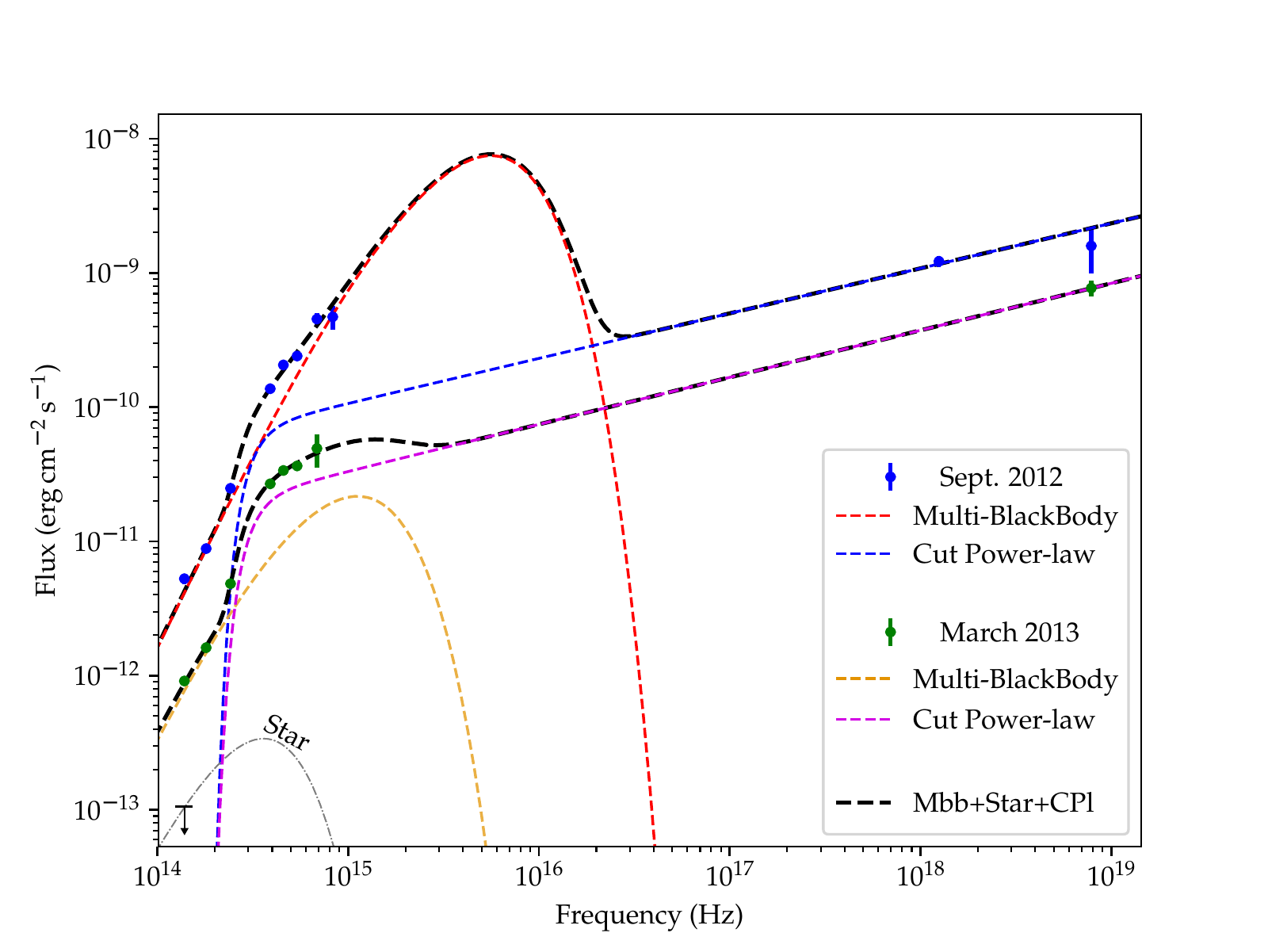}
\includegraphics[width=0.54\textwidth]{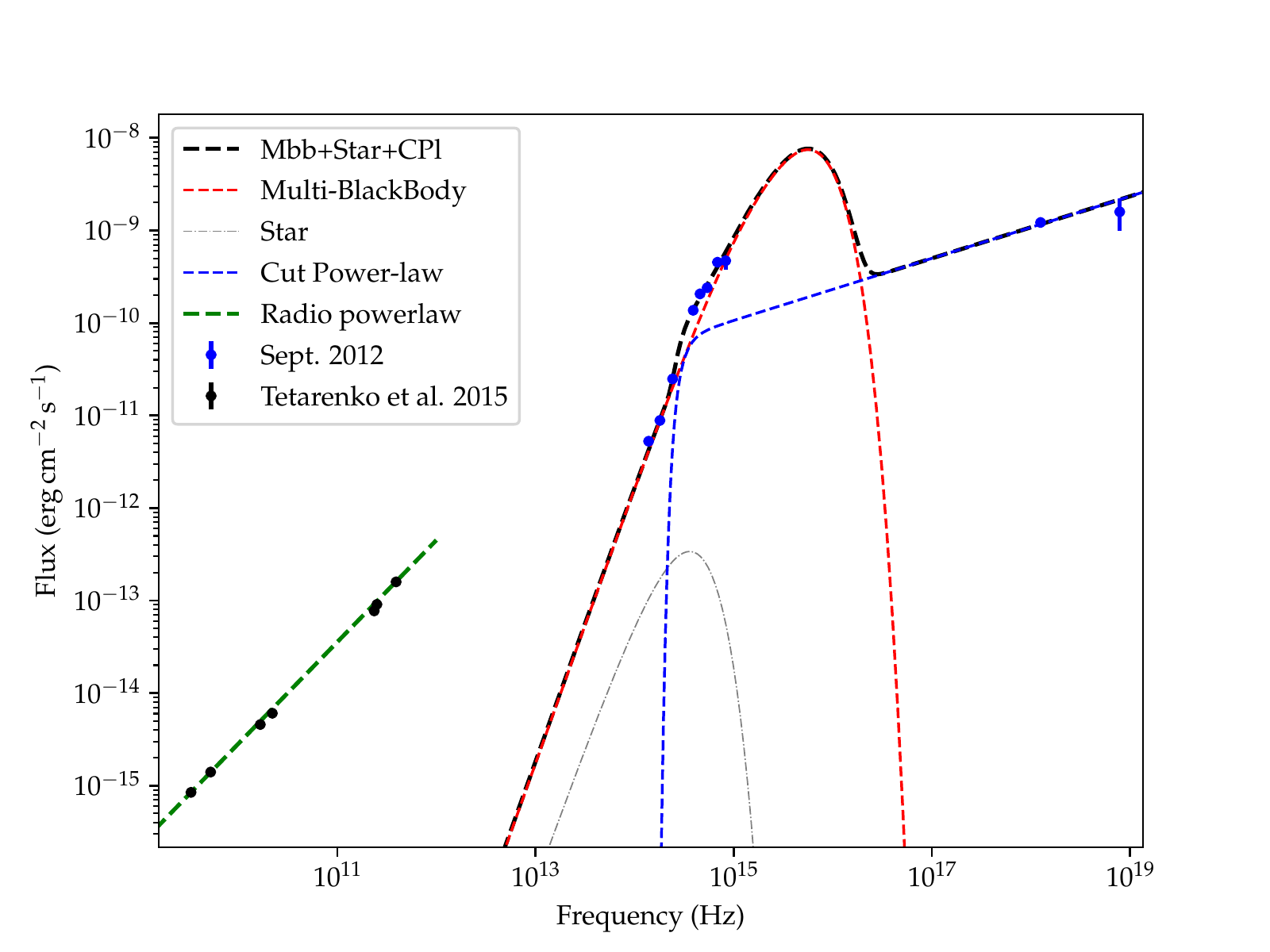}
\caption{Top panel: SED of $\swiftjdsqc$ from OIR to X-rays, for September 2012 (blue points) and March 2013 (green points) observations. The SED is built with data points de-reddened with $\Av = 7.69$. The result of the fit (plotted in black) is the sum of: multi-colour black body accretion disc (in red with R$_{in} = 1600$\,ISCO for Sept. 2012, and in yellow with R$_{in} = 2900$\,ISCO for March 2013, respectively), X-ray power law (blue for Sept. 2012 and pink for March 2013 resp.), and potentially stellar contribution (grey upper limit).
Bottom panel: Broad-band SED of $\swiftjdsqc$ from radio to X-rays for September 2012 observations, including our best fit obtained by adjusting only infrared, optical to X-ray data points (with same colour coding and R$_{in} = 1600$\,ISCO). For comprehensiveness, we also include radio data points reported in \cite{tetarenko:2015}. \label{fig:sedfit}}
\end{figure}

\subsection{Fitting September 2012 SED}

We are now able to fit the SED from September 2012, with the sum of a multi-colour black body and an X-ray power law. Initially, we performed multiple fits with parameters T$_v$ and T$_i$ free, but we obtained too much degeneracy with high uncertainties in the parameters, and a value of T$_v$ that converged towards 0, indicative of a lack of observational constraint in the UV--X-ray range. We thus consider a disc dominated either by T$_v$ or by T$_i$. Subsequently, we froze T$_v = 0$ and set T$_i$ free, with the parameter R$_{in}$ fixed within the range $1 - 30 000$\,ISCO to manually explore the parameter space. We find that we can achieve a best fit for R$_{in} = 300$\,ISCO and T$_i = 16400 \pm 700 \K$. Finally, we fixed T$_{\rm i} = 0 \K$ and allowed T$_v$ to be free, the fit converged towards R$_{in} = 1600$\,ISCO and T$_v = 13 500 \pm 600 \K$. It is degenerate, but at least the high R$_{in}$ value is consistent with the low-hard state.
Even if the lack of high-energy data could potentially allow for a lower R$_{in}$, corresponding to a high-soft state, we favour a solution with high R$_{in}$, more consistent with a low-hard state expected at the beginning of the outburst. The only way to better constrain R$_{in}$ would be to have data points within the EUV domain, which seems impossible considering the high extinction of this source \citep[such EUV data had been obtained on $\xtejodh$, a microquasar that exhibited a very low extinction, see][]{chaty:2003b}.

To summarise, we fixed five parameters \{R$_{in}$, R$_{out}$, T$_i$, i, D\} and included five free parameters \{T$_v$, A, E, $\nu_0$, P\}, with 10 data points, resulting in four degrees of freedom.
All the results of our fits are reported in Table~\ref{table:fit-results} and plotted in Figure~\ref{fig:sedfit} (top panel, with the value of R$_{in} = 1600$\,ISCO).

\subsection{Fitting March 2013 SED}

For March 2013, on the declining part of the outburst, the multi-colour black-body component contributes less to the overall emission than in September. We added a single one-temperature black body resulting from stellar emission to the disc and X-ray power law. Since, for this date, there is no contemporaneous {\textit Swift}/XRT observation, we extrapolated the single X-ray {\textit Swift}/BAT measurement to optical data points.

  We first performed multiple fits with parameters T$_v$ and T$_i$ free, but, as for September 2012, we obtain too much degeneracy in the fits, with high uncertainties in the obtained parameters, with T$_v$ close to 0, again due to a lack of observational constraint in the UV--X-ray range. Then, we froze T$_v = 0$ and allowed T$_i$ to be free, with the parameter R$_{in}$ fixed within the range $1-30000$ ISCO to manually explore the parameter space. We find that we can achieve a best fit for R$_{in} = 1300$ ISCO and T$_i = 4950 \pm 70\K$. Then, we froze T$_i = 0$ and left T$_v$ as a free parameter, the fit converges towards R$_{in} = 2900$\,ISCO, and T$_v = 4240 \pm 70\K$. It is again degenerate, but consistent with the source still being in the low-hard state, with a higher value of R$_{in}$ than in Sept. 2012, suggesting that the disc has been depleted, contributing less in OIR with the viscous temperature of the disc decreasing, consistent with the source declining towards quiescence.

In all our fits, the X-ray power law is better constrained than in September, and always converges towards the parameters reported in Table \ref{table:fit-results}, and plotted in Figure~\ref{fig:sedfit} (top panel), where we show the result for R$_{in} = 2900$\,ISCO. There is no substantial difference between 2012 Sept. and 2013 March concerning the spectral index and cutoff frequency of the X-ray power law. To summarise, as for Sept. 2012 we fixed five parameters \{R$_{in}$, R$_{out}$ ,T$_i$, i, D\} and freed five parameters \{T$_v$, A, E, $\nu_0$, P\}, with nine data points, thus three degrees of freedom.

%
%

\section{Discussion} \label{section:discussion}

In Figure~\ref{fig:sedfit} (bottom panel), we show the broad-band SED corresponding to the September 2012 observations, with our infrared-to-X-ray fit, and we added radio observations reported by \cite{tetarenko:2015}. The SED shows that the OIR contribution can be described by a viscous and/or irradiated multi-colour black-body disc model, the broad-band spectrum being compatible with the source being in the low/hard state. The maximum peak luminosity corresponds at this date to $L \sim 1.17 \times 10^{37} \ergs$ (at a distance of $3.5 \kpc$), compatible with the mentioned state.

\cite{tetarenko:2015} proposed that the power law fitting of the radio/sub-mm data could be extrapolated to shorter wavelengths, including the optical $i'$-filter observations from \cite{munoz-darias:2013}, placing a lower limit for the radio jet spectral break at either $\nu_{break} \ge 2.3 \times 10^{11}$~Hz or $\nu_{break} \ge 10^{14}$~Hz, depending on the nature of the optical emission: either completely or partially coming from the jet.

From our SED fitting, we find that the OIR component is fully consistent with the sum of thermal (optically thick) emission from the accretion disc and non-thermal (synchrotron) emission from the X-ray power law\footnote{This would naturally explain why there is no discernable feature on the IR spectrum.}. The infrared component does not seem compatible with the extrapolation of the radio power law, therefore the radio spectral break must occur at lower frequencies, meaning between $10^{12} \mbox{Hz} \le \nu_{break} \le 10^{13}$~Hz (we thus exclude the higher break frequency possibility proposed by \citeauthor{tetarenko:2015} \citeyear{tetarenko:2015}). In addition, from SED fitting we can set an X-ray spectral break at frequencies $3.1 \leq \nu_{break} \leq 3.5 \times 10^{14}$\,Hz.


For comprehensiveness, we tried another method, constraining the disc temperature parameters, by taking the L$_X$ value given by the {\textit Swift} data points, and using the relationship between L$_X$ and $\frac{dM}{dt}$ given in \cite{vrtilek:1990}. By doing this we derive T$_{i} = 4000$\,K and T$_{v} = 1600$\,K, much lower than fitted T$_{i}$ and T$_{v}$ values for September 2012, reported in Table \ref{table:fit-results}. We then obtained the irradiation of the K5V star itself, by computing the stellar T$_{eff}$ using L$_X$, following the method described in \cite{ducci:2019}. We then fitted the SED using these parameters T$_{i}$, T$_{v,}$ and T$_{eff}$, the best fit always giving distances of less than 1 kpc, which correspond to a companion star of very late spectral type, typically M5V. Such a combination of M5V companion star at a distance less than 1 kpc would make $\swiftjdsqc$ an exceptionally close-by LMXB. In addition, we point out that, whatever fitting method we use, we always find a power-law cutoff within the optical domain, in agreement with the featureless FORS2 spectra.

\begin{figure}
\centering
\resizebox{1.\hsize}{!}{\includegraphics{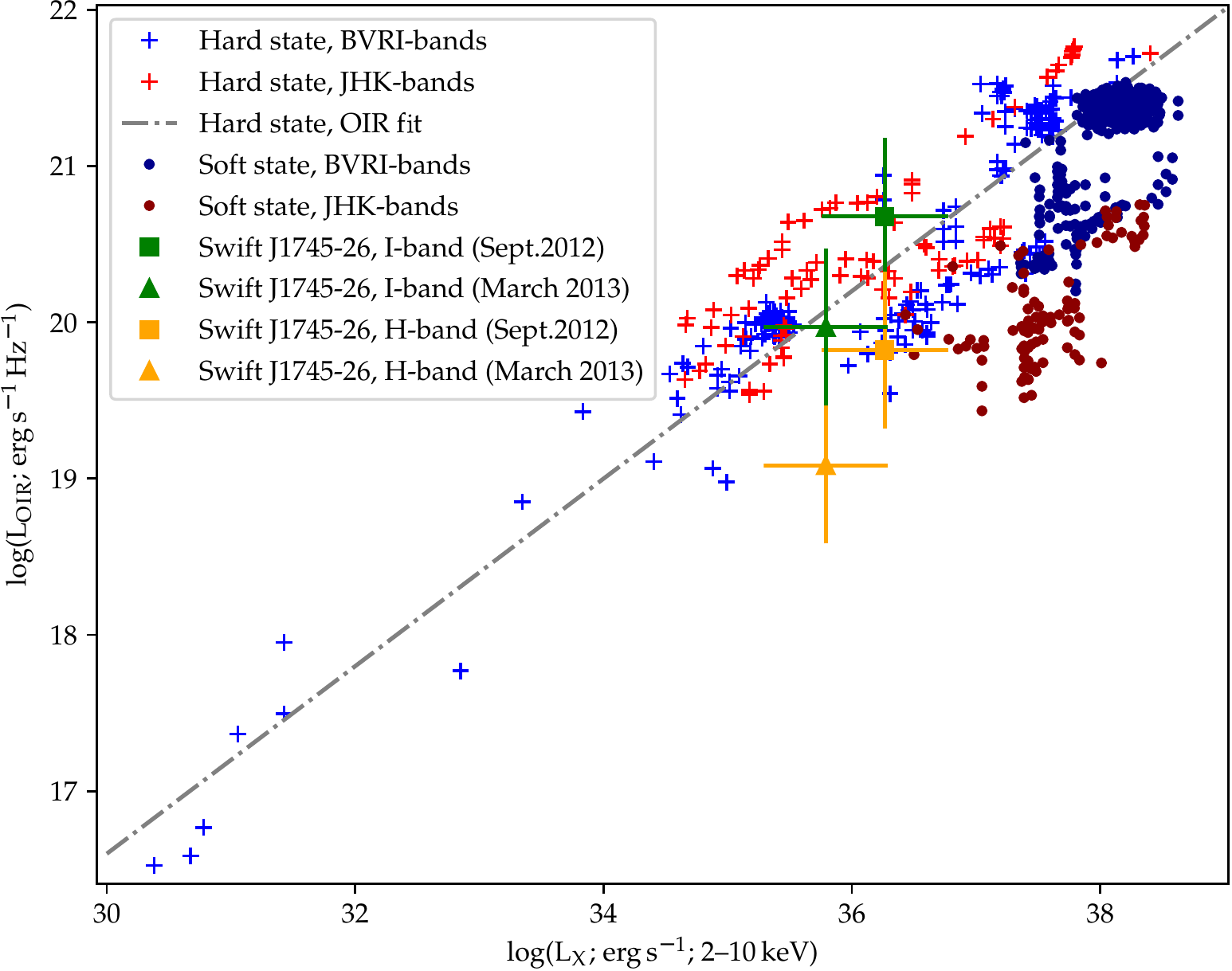}}
 \caption{OIR--X-ray luminosity diagram for black-hole LMXB in hard and soft states: data taken from \cite{russell:2006a}. $\swiftjdsqc$ luminosities, overplotted for I and H bands at $3.5 \kpc$ (green and yellow markers, respectively), are located inside the hard-state correlation, thus showing that the source remained in the low/hard state during its outburst. \label{fig:luminosities}}
\end{figure}

\cite{russell:2006a} found a correlation between OIR and X-ray emission of black-hole LMXBs. Taking the fluxes obtained in OIR for the beginning of the outburst (MJD 56188 for optical and MJD 56190 for infrared wavelengths), and X-rays \citep{tomsick:2012c, sbarufatti:2012}, we computed the luminosity of $\swiftjdsqc$ at $3.5 \kpc$ in the diagram of black-hole LMXBs (see Figure~\ref{fig:luminosities}). OIR and X-ray observations of $\swiftjdsqc$ clearly show that the source always remained in the low/hard state in both epochs, maybe entering a hard/intermediate state towards quiescence when the disc retreats, but always excluding a high-soft state (although we can not exclude a brief intrusion into this state). This confirms our results obtained from SED fitting, with a substantial contribution from radio and X-ray power laws, the OIR SED being due to a reprocessing of X-rays in a cool accretion disc. We thus suggest, in agreement with \cite{curran:2014}, that $\swiftjdsqc$ experienced during 2012-2013 a failed transition between hard and soft state, characterized by a radio jet that never completely quenched, and an accretion disc that never reached a high-soft state emission. This failed transition has already been observed in other black-hole transients, such as the microquasars $\swiftjdsct$ \citep{cadolle-bel:2007}, $\hdsqt$ \citep{capitanio:2009, chaty:2015a}, MAXI~J1836-194 \citep{ferrigno:2012} and $\xtejqcq$ \citep{curran:2013}. For instance, in the case of $\hdsqt$, optical/infrared observations showed that the black hole was both radio quiet and infrared dim in the low/hard state \citep{chaty:2015a}.

\section{Summary} \label{section:conclusion}

We performed OIR observations of the microquasar $\swiftjdsqc$ during its 2012-2013 outburst with the VLT. Our results can be summarised as follows:

\begin{enumerate}
\item During its outburst, the $\swiftjdsqc$ SED from infrared up to X-rays can be adjusted by the sum of both a viscous irradiated multi-colour black body emitted by an accretion disc, and a synchrotron power law at high energy.
\item The OIR emission of $\swiftjdsqc$ arises mainly from a multi-colour black-body accretion disc, with a power law contribution from the X-rays, similarly to $\hdsqt$ \citep{chaty:2015a}. On one hand, since the radio jet does not contribute much to the OIR emission, a spectral break at radio/sub-mm frequencies must exist at frequencies between $10^{12} \mbox{Hz} \le \nu_{break} \le 10^{13}$Hz. On the other hand, from SED fitting, we can set an X-ray spectral break at frequencies $3.1 \leq \nu_{break} \leq 3.4 \times 10^{14}$\,Hz.
  \item Our SED fitting suggests that the transient source remained in the low/hard state during its outburst, without entering the soft state (so-called failed transition). This is also consistent with the localisation of $\swiftjdsqc$ on the optical-infrared X-ray luminosity diagram.
  \item From SED fitting, we also show that the system is compatible with an absorption of $\Av \sim 7.69 \mags$ lies within a distance interval of D$ \sim[2.6-4.8] \kpc$, with an upper limit of orbital period P$_{orb} \leq 11.3$~hours, and that the companion is a spectral late-type star in the range K0 -- M0\,V. These results support the classification of $\swiftjdsqc$ as an LMXB.

\end{enumerate}

{\bf The results obtained thanks to this work show the importance of getting simultaneous and broad-band multi-wavelength observations of soft X-ray transients, from their outburst to their quiescence.}

\section{Acknowledgments}

The authors are grateful to Dave Russell for the OIR-X-ray data of black hole LMXB kindly sent to build the luminosity diagram in hard and soft state \citep{russell:2006a}. We heartfully thank Alexis Coleiro and Federico Garc\'ia for fruitful discussions, Douglas Marshall for a careful rereading of the manuscript, an anonymous referee and the A\&A editor Sergio Campana, who gave us valuable comments. This work was supported by the Centre National d'Etudes Spatiales (CNES), based on observations obtained with MINE --Multi-wavelength INTEGRAL NEtwork--, and through the post-doctoral grant awarded to A. L\'opez-Oramas. SC is also grateful to the LabEx UnivEarthS for the funding of Interface project ``Galactic binaries towards merging''. This research has made use of NASA's Astrophysics Data System Bibliographic Services and of the SIMBAD database, operated at CDS, Strasbourg, France.


\end{document}